\documentclass[amsmath,amssymb,prb,reprint]{revtex4-2}

\usepackage{graphicx}
\usepackage{dcolumn}
\usepackage{bm}
\usepackage[whole]{bxcjkjatype}

\usepackage{mathrsfs}
\usepackage{multirow}
\usepackage{amssymb}
\usepackage{amsmath}
\usepackage{mathrsfs}
\usepackage{comment}
\usepackage{braket}
\usepackage{listings}
\usepackage{here}
\usepackage{color}
\usepackage{physics}
\usepackage{ulem}

\newcommand\jeff{\ensuremath{j_\mathrm{eff}}}
\newcommand\jzeff{\ensuremath{j_{\mathrm{eff}}^z}}

\begin{document}

\title{
Multipolar ordering in the three-orbital Hubbard model
}

\author{Naoya Chikano}
\affiliation{Department of Physics, Saitama University, Saitama 338-8570, Japan}

\author{Shintaro Hoshino}
\affiliation{Department of Physics, Saitama University, Saitama 338-8570, Japan}

\author{Hiroshi Shinaoka}
\affiliation{Department of Physics, Saitama University, Saitama 338-8570, Japan}

\date{\today} 

\begin{abstract}
The ground-state phase diagrams of the three-orbital $t_\mathrm{ 2 g }$ Hubbard model are studied using a Hartree-Fock approximation.
First, a complete set of multipolar order parameters for $t_\mathrm{ 2 g }$ models defined in terms of the effective total angular momentum $\jeff$ are theoretically derived.
These order parameters can classify off-diagonal orders between $\jeff$ = 1/2 and $\jeff$ = 3/2 manifolds.
Second, through extensive Hartree-Fock calculations, the ground-state phase diagrams in the space of (1) the onsite Coulomb repulsion $U$, (2) the spin-orbit coupling (SOC) $\lambda$, and (3) the number of electrons $n$ are mapped out.
A variety of nontrivial quantum phases with \jeff--diagonal and \jeff--off-diagonal multipole orders are found.
Finally, future studies using more numerically expensive methods, such as dynamical mean-field theory are discussed.
\end{abstract}

\maketitle

\section{Introduction}
Novel phenomena arising from competing spin-orbit coupling (SOC) and electronic correlation are matters of considerable interest in condensed matter physics.
In particular, $5d$ transition metal oxide 
provides a suitable playground to study such phenomena because these interactions are comparable.

$5d$ transition metal oxides exhibit a variety of exotic quantum phases.
For example, the layered iridate Sr$_2$IrO$_4$ with a $t_\mathrm{ 2 g }^5$ ($n$=5) configuration exhibits a nontrivial spin-orbital-entangled Mott insulator\cite{Shimura1995, Cao1998, Kim2008Sr}.
The pyrochlore oxide Cd$_2$Re$_2$O$_7$, with a $t_\mathrm{ 2 g }^2$ configuration, exhibits various multipolar ordered states~\cite{Harter2017}, superconductivity at ambient pressure, and under high pressures~\cite{Hanawa2001, kobayashi2011}.
For $n = 3$, Cd$_2$Os$_2$O$_7$ shows a finite-temperature metal-insulator transition into a non-collinear magnetically ordered state~\cite{SLEIGHT1974357, Mandrus2001, Yamaura2012}. 
The emergence of novel magnetic states has been reported in various materials such as Eu$_2$Ir$_2$O$_7$~\cite{Wang2017Eu2,Chun2018Eu}, Ba$_2$YIrO$_6$~\cite{Fuchs2018Ba2YIrO6}, and Na$_2$IrO$_3$~\cite{Iro2010Na2, Liu2011Na2}.

On the theoretical side, the $t_\mathrm{ 2 g }$ three-orbital Hubbard model with SOC is the simplest model for clarifying the novel phenomena, arising from a competing SOC and electron interaction.
Recently, Sato \textit{et al.}~\cite{Sato2015a, Aaram2017, Sato2019} analyzed a model with infinite spatial dimensions for $n=4$ and $5$ using the dynamical mean-field theory (DMFT), which can accurately describe local strong correlation effects~\cite{Anisimov_1997}.
They showed that this model hosts many intriguing quantum phases, such as a spin-orbital-entangled Mott insulating phase at $n=5$~\cite{Sato2015a} and a nonmagnetic excitonic insulator~\cite{Aaram2017} with quadrupole ordering at $n=4$~\cite{Sato2019}.

However, further exploration of the phase diagram at arbitrary fillings is computationally expensive.
In particular, the hybridization-expansion continuous-time quantum Monte Carlo method ~\cite{Werner:2006iz,Werner:2006ko}
, which was used to solve the impurity model in previous studies, suffers from a severe sign problem when the SOC is strong~\cite{Kim2020Valenti,ShinaokaPyrochloreReview}.
Although efforts to alleviate the sign problem in CT-QMC continue~\cite{Kim2020Valenti}, it is still difficult to compute the low-$T$ properties of a multi-orbital Hubbard model under SOC, especially in cases which are slightly away from filling~\cite{Kim2020Valenti}.
Thus, it is important to map the phase diagram using a simpler and computationally feasible method.

Another theoretical issue is the complete classification of the (local) multipolar order parameters.
Multipole representation is suitable for the description of order parameters in spin-orbit entangled systems as developed in $f$-electron systems \cite{Ohkawa1993, Shiina1997, Kusunose2008jpsj, Kusunose2020complete}.
Although a complete clarification should involve 36 (= $6\times 6$) distinct order parameters for $t_\mathrm{ 2 g }$ systems, 
only a subset of order parameters have been used to classify quantum phases in previous studies~\cite{Sato2019}.
Moreover, these order parameters do not correctly detect the off-diagonal orders of the so-called $\jeff = 1/2$ and $\jeff = 3/2$ manifolds.

In this study, 
a complete set of multipolar order parameters designated for classifying spin-orbital entangled states, similar to Ref.~\cite{Kusunose2020complete}, are derived.
In particular, the diagonal and off-diagonal matrix elements with respect to $\jeff$ are focused upon.
These order parameters can distinguish diagonal and off-diagonal orders correctly.
Based on this result, the ground-state phase diagram of the $t_\mathrm{ 2 g }$ Hubbard model using the unrestricted Hartree-Fock approximation is systematically explored and all possible (particle-conserving) local symmetry-breaking patterns are considered.
The phase diagram in the space of three parameters: the onsite intra-orbital Coulomb repulsion $U$, the strength of the SOC $\lambda$, and the number of electrons $n$, are mapped out, thus revealing the presence of various spin-orbital entangled states and quantum multi-critical points. Furthermore, the intensity maps of multipolar order parameters as a function of $U$ and $\lambda$, which reveal several
interesting features about the non-trivial quantum phases, are studied. 

The remainder of this paper is organized as follows.
In Secs.~\ref{sec:model} and \ref{sec:method}, the model and numerical method are introduced.
In Sec.~\ref{sec:orderparam}, the multipole expansion described in a complete basis set is defined.
The relation between the order parameters introduced in this study and the conventional ones used in previous studies are also discussed. The phase diagram computed for $n=4$ is represented in Sec. \ref{sec_n4}. In Sec.~\ref{sec:general-n}, cases of general electronic filling and intensity maps of the order parameters are discussed. 
In Sec.~\ref{sec:summary}, this paper is summarized.

\section{Model}\label{sec:model}

Here, a three-orbital $t_{2{\rm g}}$ Hubbard model is considered, whose Hamiltonian is given by
\begin{align}
\mathcal{H} &= \mathcal{H}_0 + \mathcal{H}_\mathrm{int} + \mathcal{H}_\mathrm{SOC},\label{eq:ham}
\end{align}
where $\mathcal{H}_0$ is the non-interacting part, $\mathcal{H}_\mathrm{int}$ is the on-site interaction part, and $\mathcal{H}_\mathrm{SOC}$ is the spin-orbit coupling (SOC).
The non-interacting Hamiltonian $\mathcal{H}_0$ corresponds to an
orbital-diagonal semi-circular DOS with full bandwidth $W=4t$, that is, $D(\omega ) = \sqrt{4t^2 - \omega^2}/\pi$.
 $t=1$ is taken as the unit of energy.

For the interaction term, the standard, fully rotationally invariant Slater-Kanamori interactions are taken:
\begin{align}
\mathcal{H}_{{\rm int}} &=  U\sum_{i\alpha }n_{i\alpha \uparrow}n_{i\alpha \downarrow} + \sum_{i, \alpha > \beta ,\sigma }U' n_{i \alpha \sigma}n_{i \beta \sigma '} \nonumber \\
&\hspace{4mm} + \sum_{i, \alpha > \beta ,\sigma }(U' - J_{{\rm H}}) n_{i \alpha \sigma}n_{i \beta \sigma} \nonumber \\
&\hspace{4 mm} + J_{{\rm H}}\sum_{i, \alpha \neq \beta }( c_{i\alpha \downarrow }^{\dagger} c_{i\beta \uparrow } ^{\dagger} c_{i\alpha \uparrow } c_{i\beta \downarrow }  +  c_{i\alpha \uparrow }^{\dagger} c_{i\alpha \downarrow } ^{\dagger} c_{i\beta  \uparrow } c_{i\beta \downarrow } )\nonumber \\
&= \frac{1}{2}\sum_{\alpha \beta \alpha ' \beta ' \sigma \sigma '} U_{\alpha \beta \alpha ' \beta '}c_{i\alpha \sigma }^{\dagger} c_{i\beta \sigma ' } ^{\dagger} c_{i\beta ' \sigma ' } c_{i\alpha ' \sigma },
\end{align}
where $i$ is the site index, $U$ is the intra-orbital Coulomb repulsion, $U'(=U-2J_{{\rm H}})$ is the inter-orbital Coulomb repulsion, and $J_{{\rm H}}$ is Hund's coupling.
Here, $\alpha(\beta)$ and $\sigma (\sigma')$ are the orbital and spin indices, respectively. The site index $i$ in the following section is omitted for simplicity. 
The Coulomb tensor is defined as
$U_{\alpha \alpha \alpha \alpha } = U, U_{\alpha \beta \alpha \beta } = U -2J_{{\rm H}}, U_{\alpha \beta \beta \alpha  } = U_{\alpha  \alpha \beta \beta } = J_{{\rm H}} (\alpha \neq \beta)$.
Throughout the present study, $J_{{\rm H}} = 0.15 U$ is taken
which is motivated by a first-principles estimate for a typical $5d$ compound Na$_2$ IrO$_3$ ($U = 2.72$ eV, $J_\mathrm{H}= 0.23$ eV)\cite{Yamaji:2014be}.

The SOC term is written as
\begin{align}
\mathcal{H}_{{\rm SOC}} = \lambda \sum_{\alpha,\beta,\sigma ,\sigma' } \expval{\alpha| \hat{\bm{l}} |\beta}\cdot \expval{\sigma| \hat{\bm{s}} |\sigma ' } c_{\alpha \sigma }^{\dagger} c_{\beta \sigma' }  
\end{align}
where $\lambda$ is the SOC strength, and $\hat{\bm{l}}$ and $\hat{\bm{s}}$ are the angular and spin momenta, respectively. By choosing the index on the order of $xy, yz, zx$, the matrix elements of $\hat{\bm{l}}$ in the $t_{2{\rm g}}$ system and $\hat{\bm{s}}$ are given by
\begin{align}
&l_x = 
 \begin{pmatrix}
 0& 0&-i  \\
0& 0& 0  \\
i& 0& 0 \\
\end{pmatrix},\\
&l_y = 
\begin{pmatrix}
 0& i& 0 \\
-i& 0& 0  \\
0& 0& 0  \\
\end{pmatrix},\\
&l_z = 
\begin{pmatrix}
 0& 0& 0  \\
0& 0&  i  \\
0&-i&0  \\
\end{pmatrix},\\
&s_x = \frac{1}{2} 
\begin{pmatrix}
 0&  1 \\
1&   0  \\
\end{pmatrix},\\
&s_y = \frac{1}{2} 
\begin{pmatrix}
 0&  -i \\
i&   0  \\
\end{pmatrix},\\
&s_z = \frac{1}{2} 
\begin{pmatrix}
 1&  0 \\
0&   -1  \\
\end{pmatrix}.
\end{align}

Under the SOC, the effective total angular momentum $\hat{\bm{J}}_{{\rm eff}} = -\hat{\bm{L}} + \hat{\bm{S}}$ is a good quantum number for the local Hamiltonian.
Here, $\hat{\bm{L}}$ ($\hat{\bm{S}}$) is the angular (spin) momentum in the spin-orbital space and is expressed as
\begin{align}
&\hat{\bm{L}} = \sum_{\alpha,\beta, \sigma} \expval{\alpha| \hat{\bm{l}} |\beta} c_{\alpha \sigma }^{\dagger} c_{\beta \sigma }, \\
&\hat{\bm{S}} = \sum_{\alpha,\sigma, \sigma'} \expval{\sigma| \hat{\bm{s}} |\sigma ' } c_{\alpha \sigma }^{\dagger} c_{\alpha \sigma' }.  
\end{align}

 The single-particle eigenstates of 
$\hat{\bm J}_{\rm eff}^2$ and $\hat{J}_{\rm eff}^z$ 
forms a complete local $\jeff$ basis set as
 $|\jeff, \jzeff\rangle = (
|\tfrac 3 2, \tfrac 3 2\rangle, 
|\tfrac 3 2, \tfrac 1 2\rangle, 
|\tfrac 3 2, -\tfrac 1 2\rangle, 
|\tfrac 3 2, -\tfrac 3 2\rangle, 
|\tfrac 1 2, \tfrac 1 2\rangle, 
|\tfrac 1 2, -\tfrac 1 2\rangle
) $.
The basis transformation matrix from the $t_{2{\rm g}}$ basis (by choosing the index on the order of $xy \uparrow, xy \downarrow , yz \uparrow, yz \downarrow , zx \uparrow, zx \downarrow$) to the $\jeff$ basis is given by 
\begin{align}
&V = 
 \begin{pmatrix}
 \frac{1}{\sqrt{3}}& 0&0 & \frac{2}{\sqrt{6}}&0 &0 \\
  0& -\frac{1}{\sqrt{3}}&0 & 0&\frac{2}{\sqrt{6}} &0 \\
   0& \frac{1}{\sqrt{3}}&\frac{1}{\sqrt{2}} & 0&\frac{1}{\sqrt{6}} &0 \\
    \frac{1}{\sqrt{3}}& 0&0 & -\frac{1}{\sqrt{6}}&0 &\frac{1}{\sqrt{2}} \\
     0& -\frac{i}{\sqrt{3}}&\frac{i}{\sqrt{2}} & 0&-\frac{i}{\sqrt{6}} &0 \\
      \frac{i}{\sqrt{3}}& 0&0 & -\frac{i}{\sqrt{6}}&0 &-\frac{i}{\sqrt{2}} \\
\end{pmatrix},
\end{align}
where each column denotes the expansion coefficients of the corresponding $\jeff$ basis function on the $t_\mathrm{ 2 g }$ basis.

\section{Method}\label{sec:method}

In this study, a model (\ref{eq:ham}) using the unrestricted Hartree-Fock approximation at zero temperature ($T$) is solved by considering all possible local mean fields.
To investigate effects of local electronic correlations, only the uniform solutions are considered.
In this approximation, the interaction term is decoupled as
\begin{align}
&\mathcal{H}_\mathrm{int} 
= \frac{1}{2}\sum_{pqrs} U_{pqrs} c^\dagger_p c_q^\dagger c_s c_r
\nonumber
\\
&\hspace{8mm} \simeq \sum_{pqrs} (U_{pqrs} - U_{pqsr})D_{pr} c^\dagger_q c_s \nonumber \\
&\hspace{8mm}-\frac{1}{2}\sum_{pqrs} (U_{pqrs} - U_{pqsr})D_{pr} D_{qs},\label{eq:MF}
\end{align}
where, composite indices ($p$, $q$, $r$, $s$) for spin orbitals,
and a $6\times 6$ density matrix $D_{pr} \equiv \expval{c^{\dagger}_p  c_r}$ were introduced.
Note that the site index $i$ in Eq.~\eqref{eq:MF} is dropped.
Self-consistent calculations using different initial density matrices are performed to obtain the lowest-energy states.

As will be seen later, the ground state of this model may be degenerated because the spin and orbital moments can be completely decoupled without SOC.
Thus a very small SOC $\lambda=10^{-5}$ is introduced to lift the degeneracy.
Furthermore, the spin moment is always aligned along the $z$ axis.

\section{Complete set of order parameters for $t_\mathrm{2g}$ systems}\label{sec:orderparam}
\begin{table*}
 \centering
  \begin{tabular}{cclllll}
\hline
\multicolumn{2}{c}{Multipoles} & \ label&\ \ \ \ \ \ rank & \ \ time-reversal 
& $j$-parity
\\ 
\hline
\multirow{6}{*}{$j_{{\rm eff}}$-diagonal}
 &\ \  $N^{\rm 3/2,even}$ &\ \ \ $a$& \  0 (monopole) & \ \ $+$ (electric) & +
 \\
 &\ \  $M_\mu^{\rm 3/2,odd}$& \ \ \ $a$& \  1 (dipole) & \ \ $-$ (magnetic) & +
\\
 &\ \  $Q_\lambda^{\rm 3/2,even}$ & \ \ \ $a$& \ 2 (quadrupole) & \ \ $+$ & +
\\
 &\ \  $T_\xi^{\rm 3/2,odd}$ & \ \ \ $a$ & \  3 (octupole) & \ \ $-$ & +
\\
 &\ \  $N^{\rm 1/2,even}$ & \ \ \ $b$& \  0 & \ \ $+$ & +
\\
 &\ \  $M_\mu^{\rm 1/2,odd}$ & \ \ \ $b$& \  1 & \ \ $-$ & +
\\
 \hline
\multirow{4}{*}{$j_{{\rm eff}}$-offdiagonal}
 &\ \  $M_\mu^{\rm offd,odd}$ & \ \ \ $c$& \  1 & \ \ $-$ & $-$
\\
 &\ \  $Q_\lambda^{\rm offd,even}$ & \ \ \ $c$& \  2 & \ \ $+$ & $-$
\\
 &\ \  $M_\mu^{\rm offd,even}$ & \ \ \ $d$& \ 1 & \ \ $+$  & $-$
\\
&\ \  $Q_\lambda^{\rm offd,odd}$ & \ \ \ $d$& \ 2 & \ \ $-$ & $-$

\\
 \hline
  \end{tabular}
 \caption{List of the complete multipole operators in three-orbital model with spin-orbit coupling.
 The suffices are $\mu = x,y,z$, $\lambda = xy, yz, zx, 3z^2-r^2,x^2-y^2$, $\xi=xyz, x(5x^2-3r^2), y(5y^2-3r^2), z(5z^2-3r^2), x(y^2-z^2), y(z^2-x^2), z(x^2-y^2)$ with $r^2=x^2+y^2+z^2$.
 }
 \label{tab:multipolemain}
\end{table*}

The symmetry-breaking information is fully encoded in the single-particle density matrix.
Multipole is a useful tool for describing the entanglement of spin and orbit degrees of freedom.
The multipole moment is described by a polynomial form of the effective total angular momentum $\hat{\bm{J}}_{{\rm eff}} = -\hat{\bm{L}} + \hat{\bm{S}}$.
Such multipoles are introduced by means of crystallographic point groups.

The multipole representation used so far for the spin-orbit-coupled three-orbital Hubbard model~\cite{Aaram2017, Sato2019} are reviewed first.
In a $t_\mathrm{2g}$ orbital system, a conventional 
multipole is constructed as a polynomial of 
$\hat{\bm J}_{\rm eff}$, and is described by 16 order parameters. The density matrix $D$ is expanded as follows:
\begin{align}
D  = \sum^{16}_{\xi = 1} C_{\xi} O_{\xi},\label{eq:rho_expand}
\end{align}
where $C$ is the weight of $O$ and $\xi$ is the index of the basis set. 
The breakdown is one order parameter(OP) for the electric monopole $N$, three OPs for the magnetic dipole $M$, five OPs for the electric quadrupole $Q$, and seven OPs for the magnetic octupole $T$. The higher-rank tensors are trivially zero. 
Operator $O$ is imposed the orthonormality
\begin{align}
{\rm Tr}\left[ O_{\xi} O_{\eta}^{\dagger} \right] = \delta_{\xi \eta}.
\end{align} 
The weight of the order parameter $C$ can be computed as
\begin{align}
C_{\xi} = {\rm Tr}\left[ D O_{\xi}^{\dagger} \right].\label{eq:trace}
\end{align} 
However, 
this definition does not produce a complete basis set because it requires $6 \times 6 = 36$ order parameters.

\begin{figure}[tbp]
  \begin{center}
   \includegraphics[width=0.5\columnwidth]{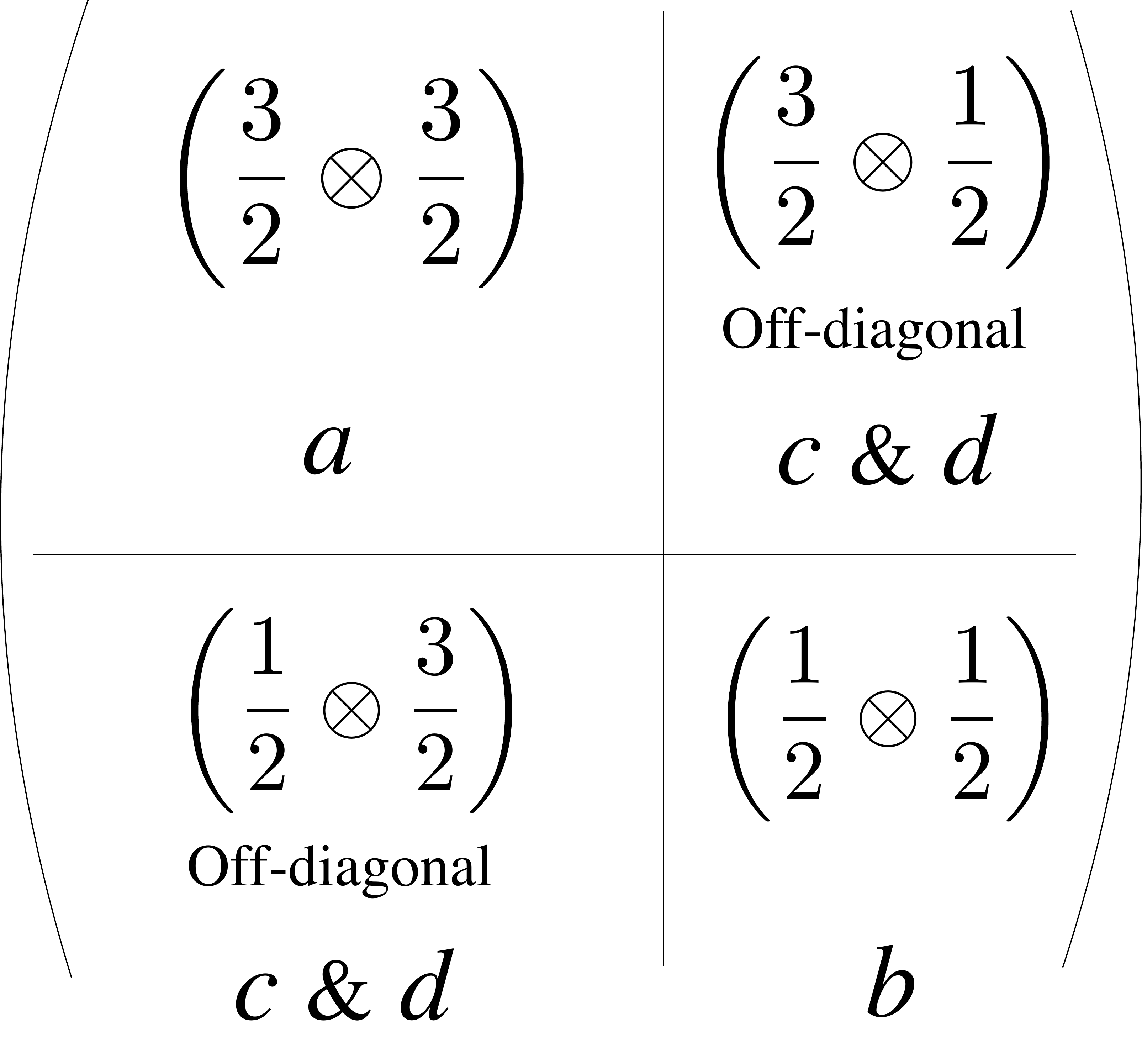}
  \end{center}
  \vspace{-2em}
  \caption{Schematic illustration of multipolar order parameters. The density matrix is decomposed into four blocks in terms of hybridization between the \jeff\,orbitals. Block $a$ has one monopole, three dipoles, five quadrupoles, and seven octupoles. Block $b$ contains one monopole and three dipoles. Blocks $c$ and $d$ have three dipoles and five quadrupoles.
  For more details, see the text.}
  \label{fig:abcd}
\end{figure}

To construct a complete set, the density matrix is decomposed into 36 one-particle tensor operators $O$ using the local projection operator. Note that multipoles for this complete set are not constructed by a polynomial of $\hat{\bm{J}}_{{\rm eff}}$, but $\bm{K} = \hat{\bm{L}} + \hat{\bm{S}}$ (see Appendix \ref{app_a}). 
which makes it possible to account for a component perpendicular to $\hat{\bm J}_{\rm eff}$. 
These tensor operators are categorized into four blocks: diagonal components $\ket{j_{{\rm eff}} = 3/2} \otimes \ket{\jeff = 3/2} $, $ \ket{1/2} \otimes \ket{1/2} $, off-diagonal components $\ket{3/2} \otimes \ket{1/2}$, and $\ket{1/2} \otimes \ket{3/2}$.
They are labelled as $a, b, c$, and $d$ in the definition order. A schematic illustration of this process is presented in Fig.\,\ref{fig:abcd}. Components $c$ and $d$ show that there is hybridization between the $\jeff = 3/2$ and $1/2$ orbitals. 
Note that the OPs in blocks $c$ and $d$ are given by a linear combination of the off-diagonal components to make it Hermitian. A time-reversal even and odd is imposed on the $c$ and $d$ components, respectively. 

The complete classification of multipoles are derived in terms of the quantum numbers of rank, time reversal, and $j$-parity.
The results are summarized in Table~\ref{tab:multipolemain} (see Appendix A for details of the classification).
The advantage of the present classification is that one can identify diagonal and off-diagonal orders in terms of \jeff~manifolds separately using the four labels ($a$, $b$, $c$, $d$).
The density matrix $D$ is decomposed into two monopoles $N$, 12 dipoles $M$, 15 quadrupoles $Q$, and seven Octupoles $T$. 
Here, the $j$-parity distinguishes the diagonal and off-diagonal components and is similar to $sp$-hybridized systems where the parity operator distinguishes the angular momentum $\ell=0$ or $\ell=1$.
In previous studies~\cite{Aaram2017, Sato2019}, they found quantum phases with off-diagonal orders, which was called ``excitonic phases''.
They identified the existence of these phases by the off-diagonal elements in the hybridization functions on the $\jeff$ basis.
The present scheme allows to directly detect and classify such phases even further.

\begin{figure}[htbp] 
  \begin{center}
  \includegraphics[width=0.9\columnwidth]{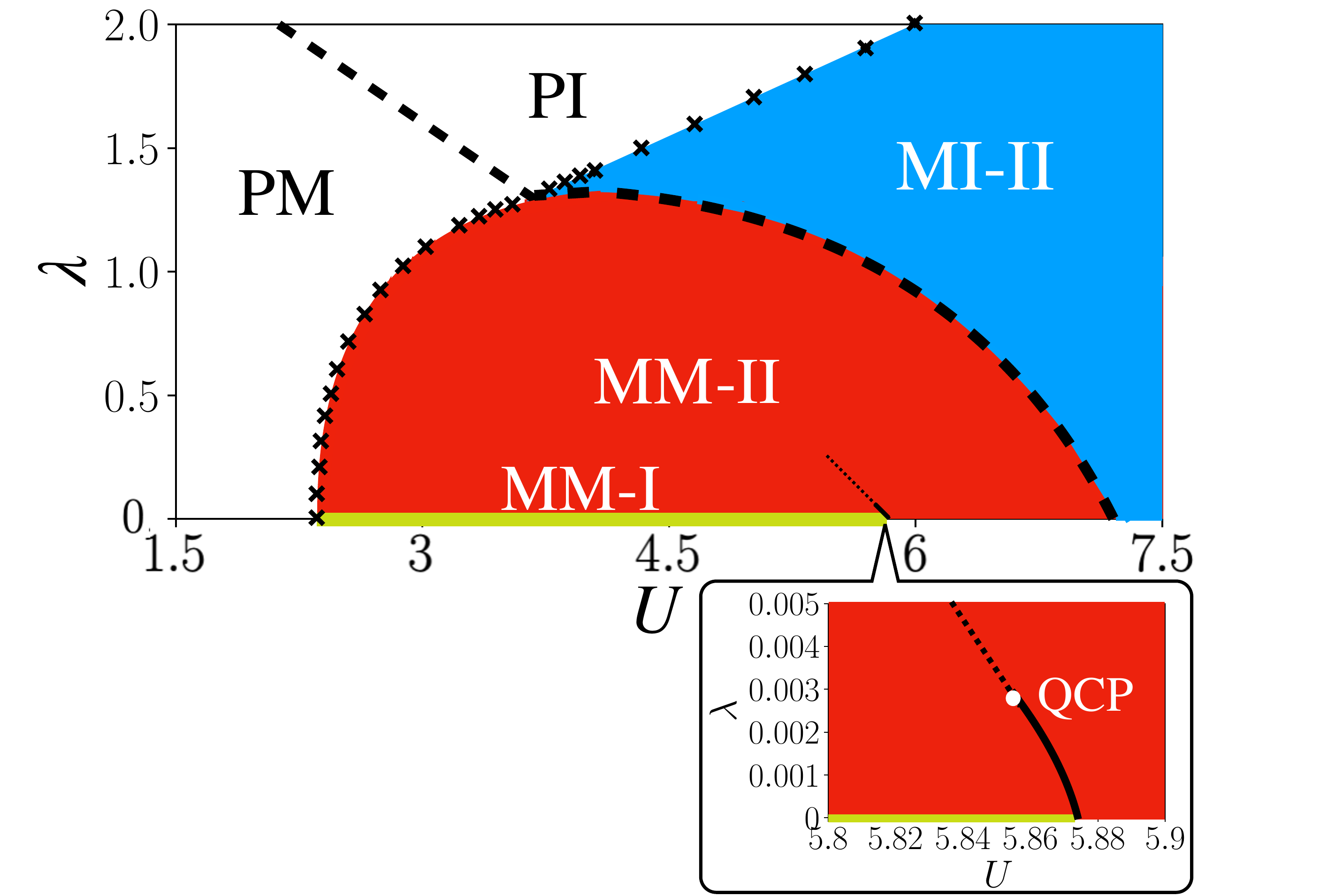}
  \end{center}
  \caption{$U$--$\lambda$ phase diagram for $n = 4$. PM, PI, MM, MI, and QCP represent the paramagnetic metal phase, paramagnetic insulator phase, magnetic metal phase, magnetic insulator phase, and quantum critical point, respectively. The numbers (I) and (II) of the MM phase denote the absence and presence of the spontaneous breaking of the orbital degeneracy, respectively.
  The dashed line denotes a metal-insulator transition.
  }
  \label{fig:n4}
\end{figure} 
\begin{figure}[htbp]
  \begin{center}
   \includegraphics[width=0.8\columnwidth]{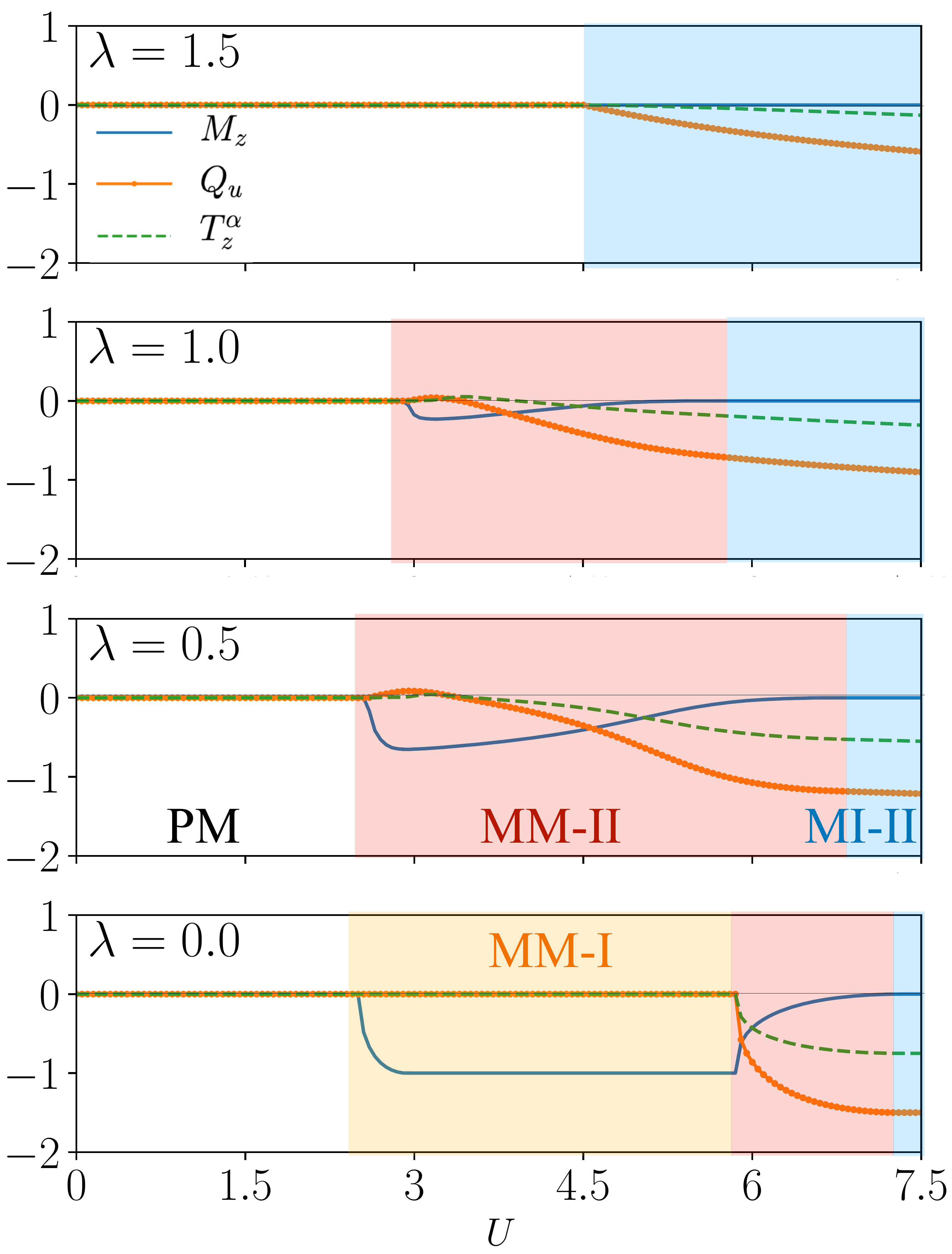}
  \end{center}
  \caption{Conventional OPs computed as a function of $U$ at $n = 4$. The background colors correspond to those of the phases shown in Figs.~\ref{fig:n4}).
  }
  \label{fig:lamb0opold}
\end{figure}
\begin{figure}[tbp]
  \begin{center}
  \includegraphics[width=0.8\columnwidth]{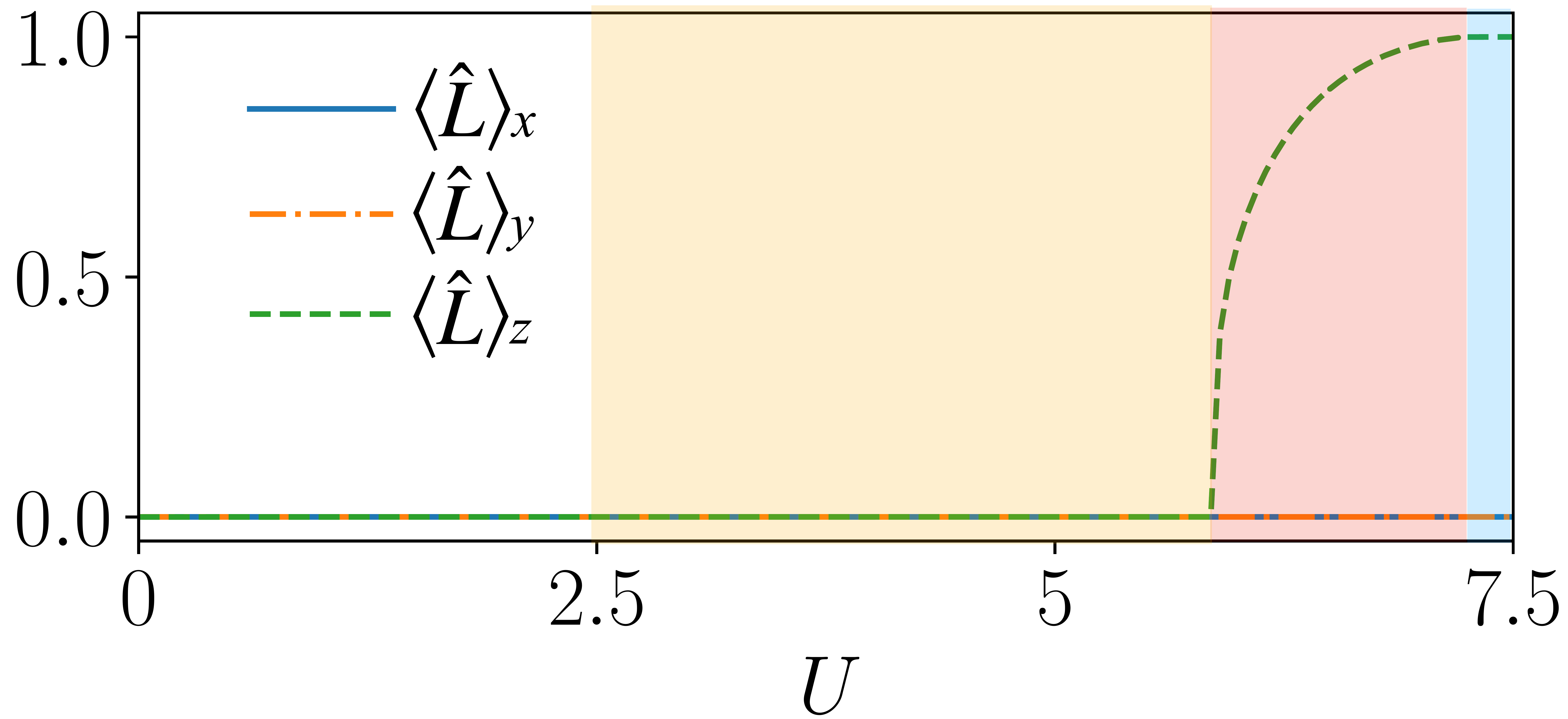}
  \end{center}
  \vspace{-2em}
  \caption{Angular momentum $\hat{\Braket{\bm{L}}}_x, \hat{\Braket{\bm{L}}}_y$ and $\hat{\Braket{\bm{L}}}_z$ computed as a function of $U$ at $n=4$ and $\lambda = 0$. The background colors correspond to those of the phases shown in Figs.~\ref{fig:n4}.}
  \label{fig:L}
\end{figure}
\section{Results for $n = 4$}\label{sec_n4}
In this section, results for $n=4$ are discussed.
Figure~\ref{fig:n4} shows a $U$--$\lambda$ phase diagram computed for $n=4$, which involves four different phases: a paramagnetic metallic (PM) phase, a magnetic metallic (MM) phase, a paramagnetic insulating (PI) phase, and a magnetic insulating (MI) phase.
The labels (I) and (II) denote the absence and presence of non-zero higher-rank multipoles $Q$ and $T$, respectively, as shown in Fig.~\ref{fig:lamb0opold}.
The detailed analysis of multipole order parameters will be given in the next subsections.
In the PM and PI phases, no spontaneous symmetry breaking takes place.
At a larger $U$, the time-reversal symmetry is broken in the MM and MI phases.
In the phase diagram, there is a dome of MM-II at moderate values of $U$, while the MI is always stable in the strong $U$ limit.
In these magnetic phases, the electric quadrupole is also active.

Note that the PI phase corresponds to the relativistic band insulator at $n = 4$, where the $\jeff = 3/2$ and $\jeff= 1/2$ bands are completely separated in energy by the strong SOC.
The critical value $\lambda_\mathrm{c}$ of the metal-insulator transition between the PM and PI is $\lambda_\mathrm{c} = 8/3$ at $U=0$.
As seen in Fig.~\ref{fig:n4}, $\lambda_\mathrm{c}$ decreases as $U$ increases until it reaches the magnetic transition line.
On the other hand, the boundary between PI and MI-II is determined by the competing $\lambda$ and $J_\mathrm{H}$, and the slope of the boundary is roughly proportional to $J_\mathrm{H}$ (not shown). 

\begin{figure}
 \begin{minipage}{\hsize}
  \begin{center}
   \includegraphics[width=\textwidth]{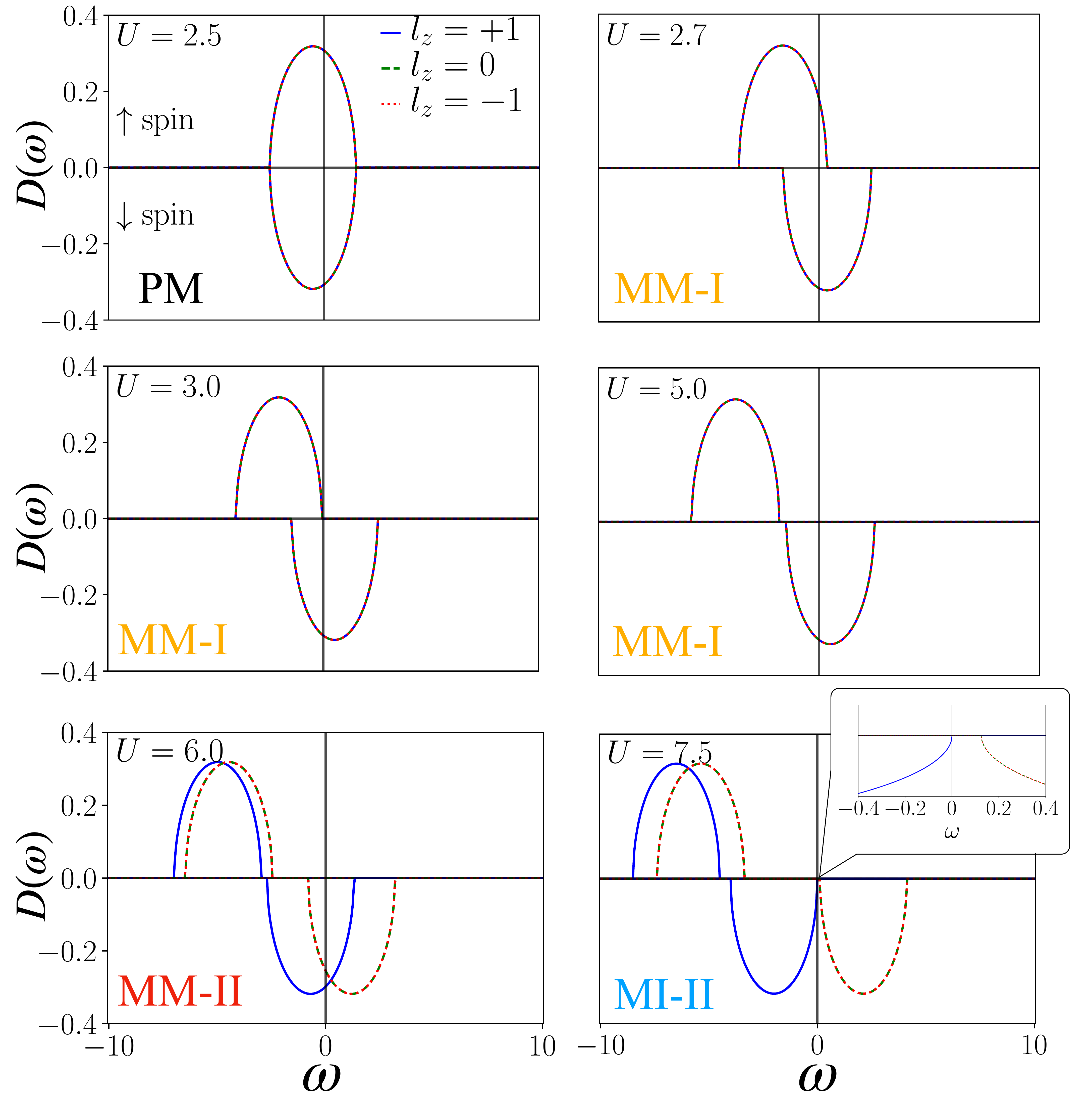}
  \end{center}
  \vspace{-2em}
  \caption{$U$ dependence of the computed DOS projected onto $l_z$=0,~$\pm 1$ orbitals computed at $n  = 4$ and $\lambda = 10^{-5}$. The Fermi energy was $\omega = 0$.
  }
  \label{fig:dosudep-n4-lambda0}
  \end{minipage}
\end{figure}

\begin{figure}[htbp]
 \begin{minipage}{\hsize}
  \begin{center}
   \includegraphics[width=\textwidth]{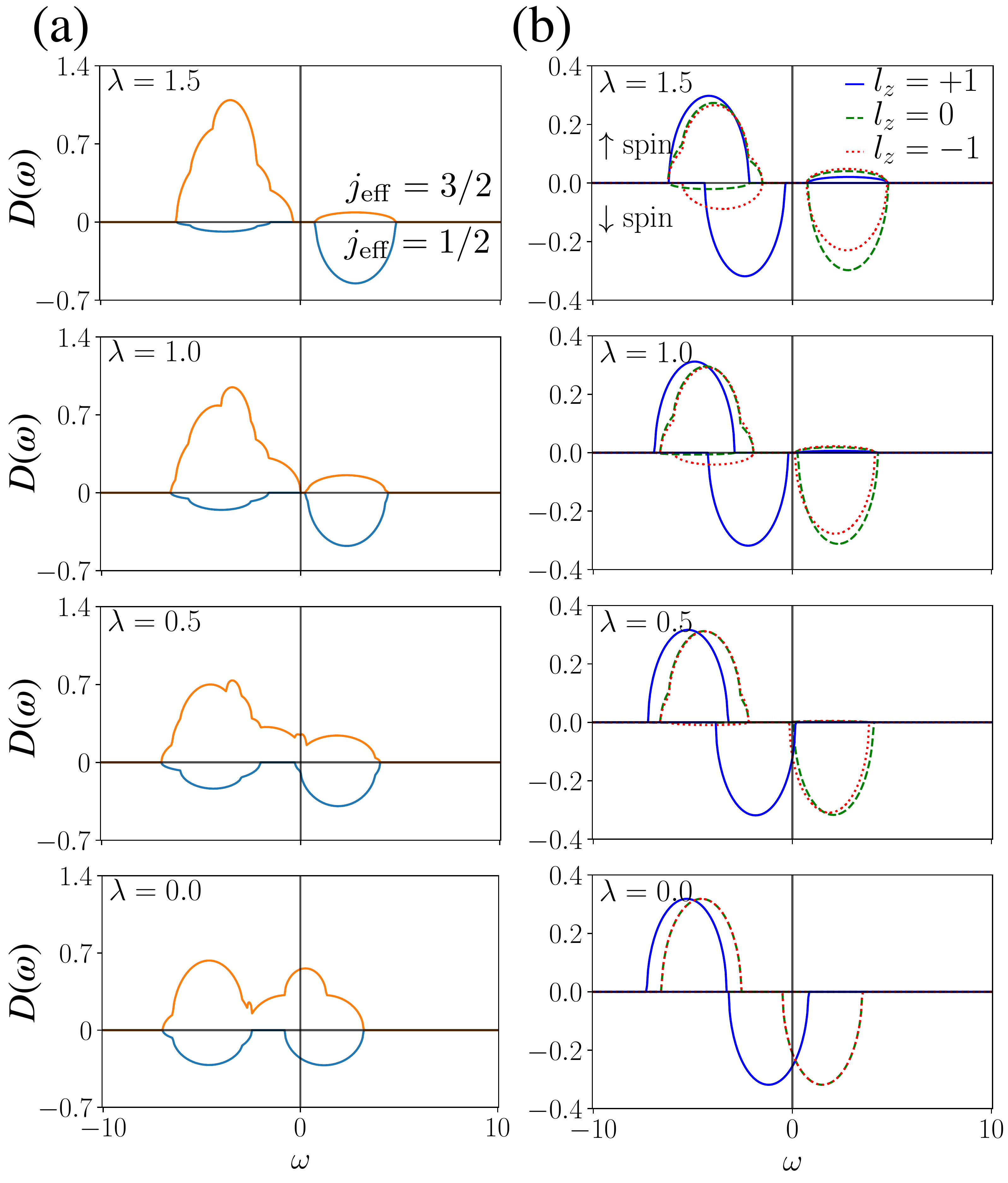}
  \end{center}
  \caption{$\lambda$ dependence of the computed partial DOS for $n  = 4$ and $U = 6$. The DOS is projected onto (a) the $\jeff$=1/2 and 3/2 orbitals and (b) $l_z$=0,~$\pm 1$ orbitals. The Fermi energy is located at $\omega = 0$.
  }
  \label{fig:dosujlz}
  \end{minipage}
\end{figure}

\subsection{Results for $\lambda=0$}
First, the results of conventional OPs computed for $\lambda=0$ are discussed. In this study, the result of $\lambda = 10^{-5}$ is treated as $\lambda = 0$ because the spin and orbital moments are completely decoupled at $\lambda=0$ in the MI phase.
As seen in Fig.~\ref{fig:lamb0opold}, 
all the conventional OPs are zero up to $U\simeq 2.6$.
Since the magnetic moment is aligned along the $z$ axis,
only $M_z, Q_u( 3z^2 - r^2)$ and $T^{\alpha}_z (z(5z^2 - 3r^2))$ are non-zero.
A transition point separates the PM phase and the MM-I phase around $U_{{\rm c}1}\simeq 2.6$.
This transition is characterized by the spontaneous breaking of the time-reversal symmetry through the emergence of $M_z\neq 0$.
This is followed by a subsequent first-order transition at $U_{{\rm c}2}\simeq 5.86$, which is characterized by the emergence of $Q_u$ and $T_z^{\alpha}$.
This corresponds to a spontaneous breaking of the orbital degeneracy.
As shown in Fig.~\ref{fig:L}, the angular momentum $L_z$ becomes finite for $U \geq U_{{\rm c}2}$.
With a further increase in $U$, the metal-insulator transition occurs, and $L_z$ reaches one at $U_{{\rm c}3} \simeq 7.2$.  
The symmetry of the system does not change across this transition point.

Figure~\ref{fig:dosudep-n4-lambda0} shows the partial density of states (DOS) projected onto the angular momentum $l_z$=$0, \pm1$ orbitals for typical values of $U$ in the PM, MM-I, and MM-II phases.
Note that the mean fields are diagonal in the $l_z$ basis for these parameters.
The six basis vectors are labelled by the spin $(\uparrow, \downarrow)$ and angular momentum $l_z$=$0 ,\pm1$. 
A magnetic transition happens at $U_{c\rm{1}} \sim 2.6$.
At $U \simeq 3$, in the MM-I phase, a Lifshitz transition occurs, and then the spin becomes fully polarized as $\hat{\Braket{\bm{S}}}_z =1$.
As seen in Fig.~\ref{fig:dosudep-n4-lambda0}, the three up-spin orbitals are completely filled, leaving one electron in the down-spin orbitals.
The partially filled three down-spin orbitals are still degenerate, and thus $\hat{\Braket{\bm{L}}}_z = 0$.
After the spontaneous breaking of the orbital degeneracy ($U \geq U_{{\rm c}2}$),
the partial $l_z$=+1$\uparrow$ DOS is separated from the partial $l_z$=$0$$\uparrow$ and partial $l_z$=$-1$$\uparrow$ DOS. 
At $U = U_{c3}$, a finite gap finally opens at the Fermi energy, and 
the system becomes a magnetic insulator with $\hat{\Braket{\bm{S}}}_z = 1$ and $ \hat{\Braket{\bm{L}}}_z= 1$.
Note that the spin and orbital moments are coupled in an antiparallel manner owing to the small SOC.
In the fully spin-polarized situation, the self-consistent equation is simplified. See Appendix B for further details.

\begin{figure}[tbp]
 \begin{minipage}{\hsize}
  \begin{center}
   \includegraphics[width=\textwidth]{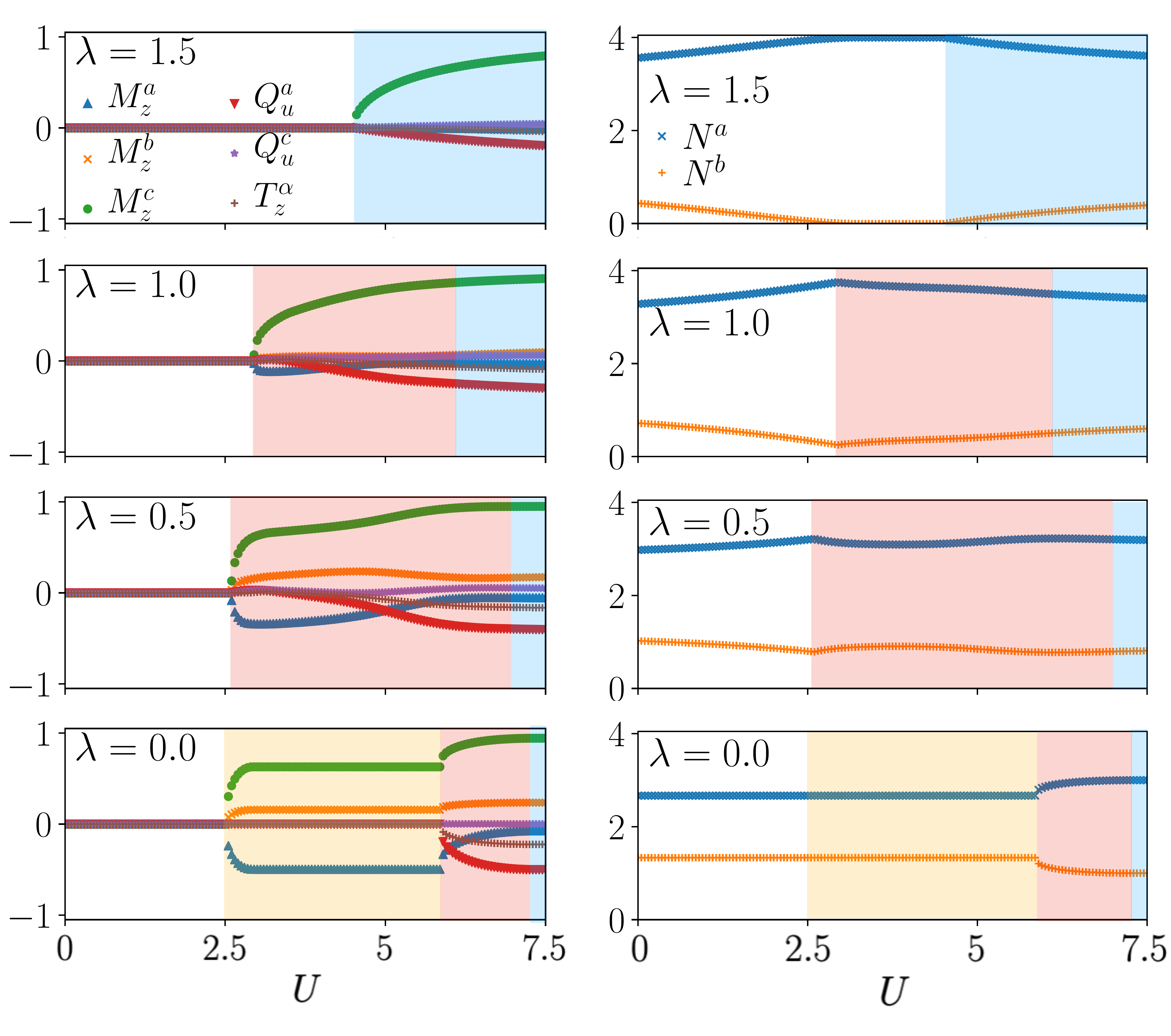}
  \end{center}
  \caption{Complete OPs computed as a function of $U$ at $n = 4$.
  The superscripts $a, b$, and $c$ denote $\jeff = 3/2$ components, $\jeff = 1/2$ components, and $\jeff = 3/2$--$1/2$ entangled components, respectively. The background colors correspond to those of the phases shown in Figs.~\ref{fig:n4}). 
  }
  \label{fig:lamb0opnew}
 \end{minipage}
\end{figure}

\subsection{Results for $\lambda>0$}
Once $\lambda$ is turned on, $Q_u$ and $T_z^{\alpha}$ become coupled to $M_z$.
Thus, 
the MM-II phase
is characterized by $M_z, Q_u, T_z^{\alpha} \neq 0$.
As shown in the inset of Fig.~\ref{fig:n4}, the first-order transition terminates at a quantum critical endpoint at $\lambda \sim 0.003$ and $U \sim 5.68$.
On the other hand,
the critical value $U$ of the magnetic transitions increases as $\lambda$ increases, and the SOC suppresses magnetization. 
Figure \ref{fig:dosujlz} shows the $\lambda$ dependence of the partial DOS projected onto the $\jeff$ basis and the $l_z$ basis at $U = 6$.
In the weak SOC limit (MM-II), the partial $l_z$ DOS is not distorted, so the $l_z$ scheme is a better representation.
As $\lambda$ is increased,
the $\jeff=1/2$ contributions become dominant for the unoccupied DOS, whereas those of $\jeff=3/2$ become dominant for the occupied DOS.
In this regime, the $\jeff$ representation is better. 
Figure~\ref{fig:lamb0opnew} shows the complete OPs computed as a function of $U$ at $n=4$. $M_z^a, M_z^b, M_z^c, Q_u^a, Q_u^c$ and $T^{\alpha a}_z$ are non-zero.
The difference between conventional (Fig.~\ref{fig:lamb0opold}), and the complete OPs are clearly seen in the magnetic insulator phase.
With increasing $U$, one of the old OPs, $M_z$, decreases toward zero and vanishes completely for $U \geq U_{c3}$.
In terms of the complete OPs, however, the off-diagonal component of the magnetic dipole $M^c_z$ remains the most dominant.

\section{Results for general filling}\label{sec:general-n}
In this section, the filling $n$ dependence of the ground-state phase diagrams are discussed.
\subsection{$n$--$U$ phase diagram at $\lambda=0$}
\begin{figure}[htbp]
  \begin{center}
  \includegraphics[width=\linewidth]{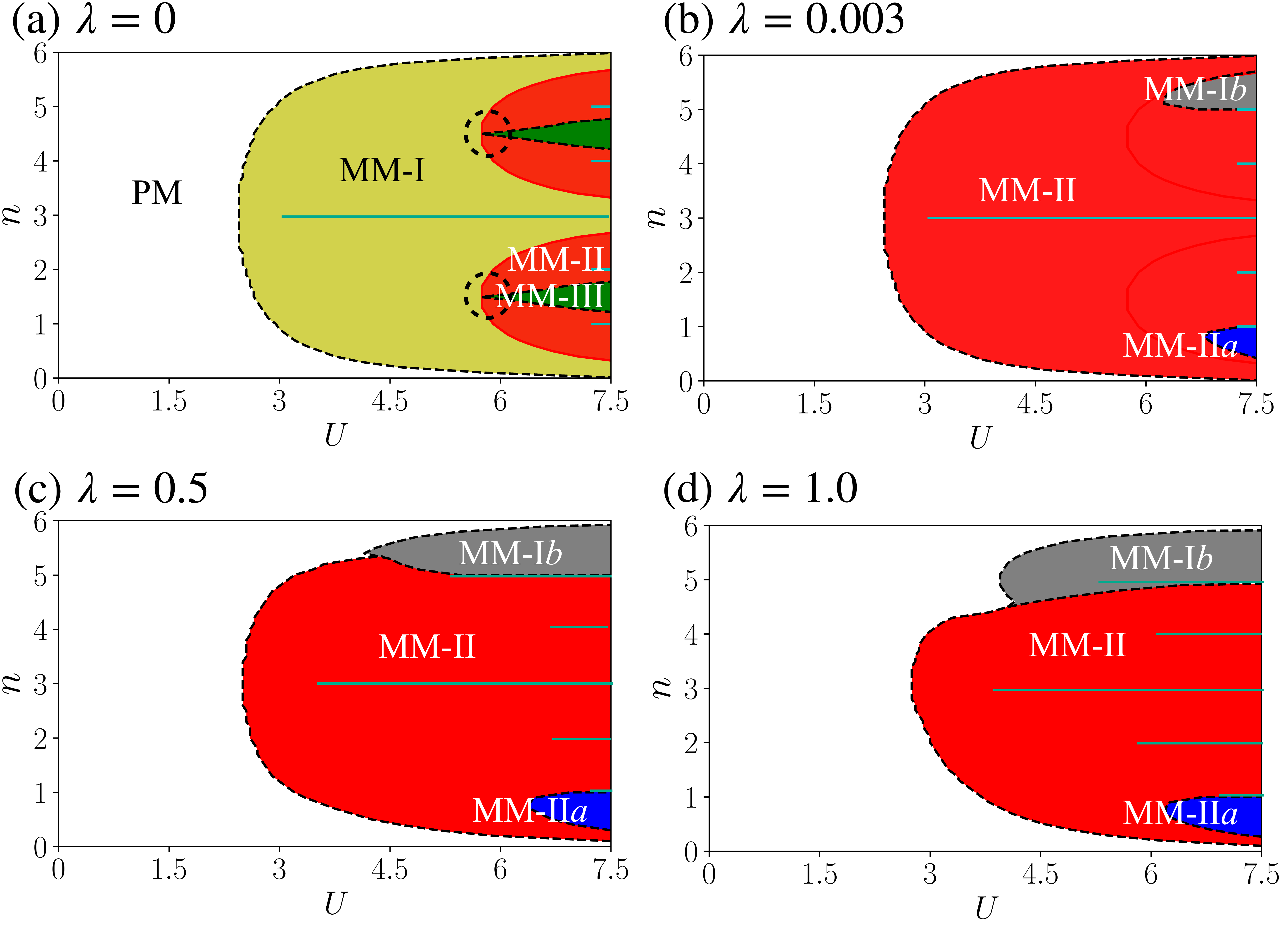}
  \end{center}
  \caption{The $n$--$U$ phase diagram for (a) $\lambda=0$, (b) $\lambda=0.003$, (c) $\lambda=0.5$ and (d) $\lambda=1.0$. PM and MM represent a paramagnetic metal phase and magnetic metal phase, respectively.
  The numbers I and II(III) indicate the absence and appearance of the higher-rank multipoles $Q$ or $T$, respectively.
  The solid and broken lines indicate the first-order and second-order transitions, respectively.
  The blue lines on $n = 1,2,3,4$, and $5$ indicate the insulating phases.
  The broken circles denote multiple critical points where the first-order transition lines merge (see text).
  }
  \label{fig:n-U-pd}
\end{figure}

\begin{figure}[htbp]
 \begin{minipage}{\hsize}
  \begin{center}
   \includegraphics[width=\linewidth]{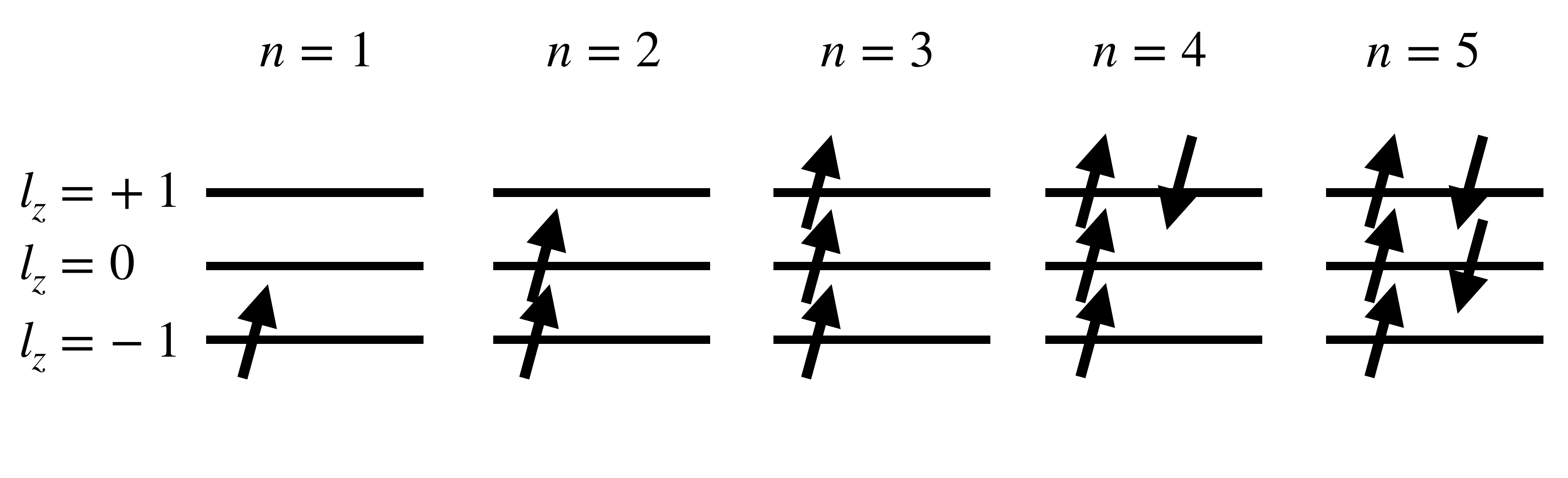}
  \end{center}
  \caption{Schematic illustration of electronic configurations projected onto $l_z$=0,~$\pm 1$ orbitals at $\lambda = 0$ and $U = 7.5$. 
  Note that the mean fields and density matrices are diagonal in this basis.
  }
  \label{fig:fill}
\end{minipage}
\end{figure}

Figure~\ref{fig:n-U-pd}(a) shows the $n$--$U$ ground-state phase diagram computed for $\lambda= 0$.
The PM exits only for a small $U$ at $\lambda=0$.
With an increment in $U$, the ground state turns into the magnetic metallic phase (MM-I), where only the magnetic dipole $M_z$ is finite.
With a further increase in $U$, the MM-I phase turns into the MM-II phase through a first-order transition.
In the MM-II phase, the symmetry of the orbital is broken, and the angular momentum $L$ and higher-order multipole order parameters are finite as
$L_z\neq 0$, $M_z\neq 0$, $Q_u\neq 0$, $T_z^\alpha \neq 0$ (see Figs.~\ref{fig:lamb0opold}, and \ref{fig:L}).
At $n=3$, symmetry breaking in the orbital sector does not occur because $\hat{\Braket{\bm{L}}}_z = 0$ and $\hat{\Braket{\bm{S}}}_z = 3/2$, as illustrated in Fig.~\ref{fig:fill}.

\begin{figure}[tbp]
  \begin{center}
   \includegraphics[width=\linewidth]{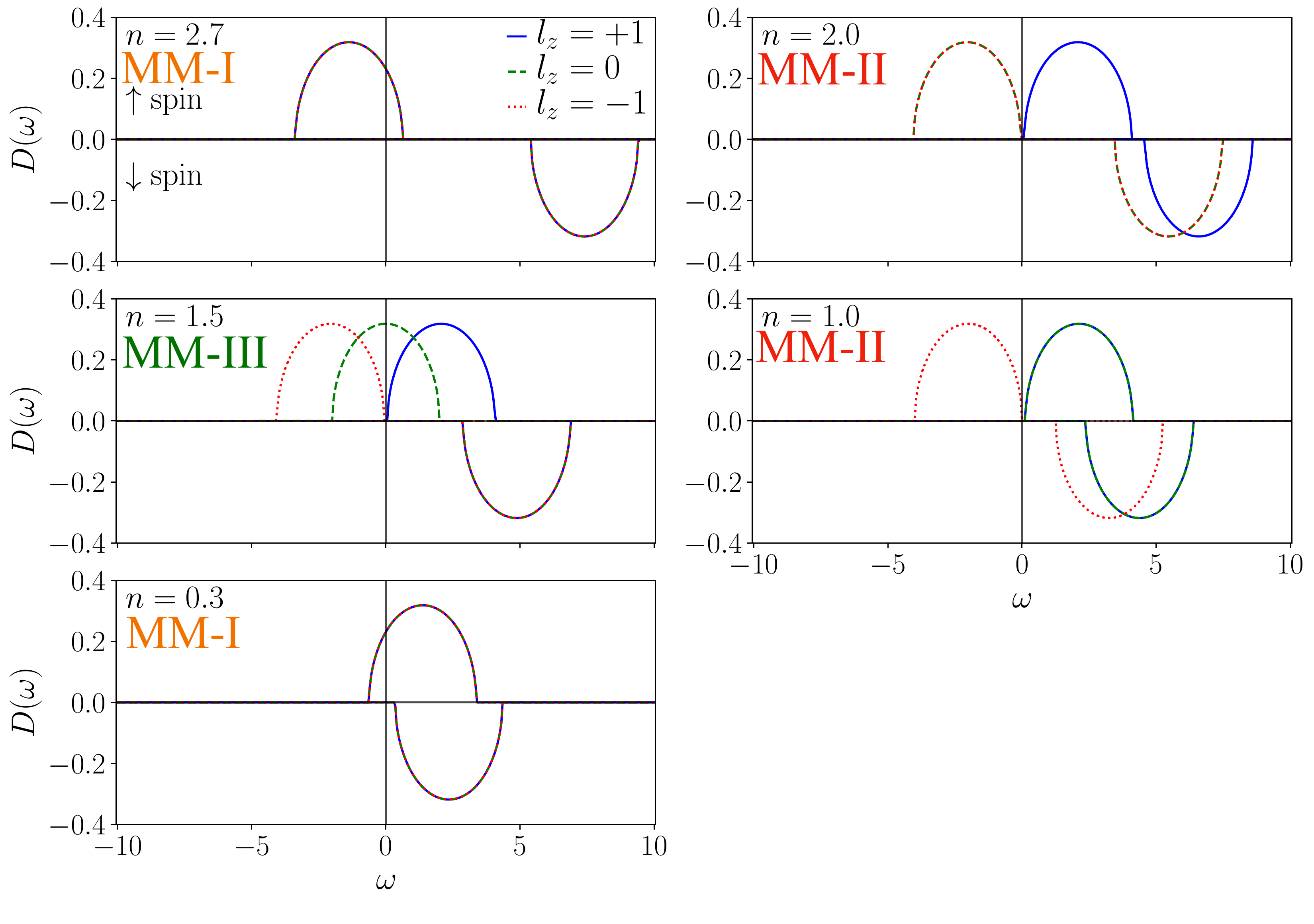}
  \end{center}
  \caption{$n$ dependence of the computed DOS projected onto $l_z$=0,~$\pm 1$ orbitals for three different phases ($U = 7.5$, $\lambda =0$). The Fermi energy is located at $\omega = 0$.}
  \label{fig:dosu75}
\end{figure}

Furthermore, a new distinct phase (MM-III) emerges only for $1 < n < 2$ and $4 < n < 5$.
The transition between MM-II and MM-III is second-order.
The three MM phases can be distinguished by different degeneracies of the DOS projected onto the $l_z$ orbitals for the major spin.
Figure~\ref{fig:dosu75} plots typical data of DOS at $U=7.5$.
In the MM-I phase, all three orbitals are degenerate for each spin.
In the MM-II phase, only two of the three orbitals remain degenerate.
In contrast, in the MM-III phase, all three orbitals become non-degenerate.
The MM-I, MM-II, and MM-III phases meet at $U \sim 5.5$, $n = 1.5$, and $4.5$.
Moreover, the first-order transitions between MM-II and MM-III become continuous only at these critical points.

The insulating states can be stable only if $n$ is an integer.
Except for $n =3$, the metal-insulator transition occurs at $U_{{\rm c}3} \sim 7.2$.
At $n = 3$, the metal-insulator transition occurs at a smaller $U \sim 3$ because a high-spin configuration is stable.
Figure~\ref{fig:fill} illustrates the electron configurations projected onto the $l_z$ orbitals in the insulting phases at $n=1,2,\cdots,5$.
At $n=3$, 4, and 5, the spin-up orbitals are fully occupied for a large $U$.
Thus, the system can be effectively regarded as a spinless system of spin-down orbitals.
This gives rise to an interesting emergent symmetry between $n=4$ and $n=5$:
A particle-hole transformation for the spin-down orbitals connects these two states. 
Consequently, the MM-II phase is symmetric with respect to $n=4.5$, as shown in Fig.~\ref{fig:n-U-pd}(a). 
This symmetry originates from the fact that the spin-up orbitals are fully occupied,
which is due to the ignorance of quantum spin fluctuations in the zero-$T$ Hartree-Fock calculations.
See Appendix~\ref{appendix:symm} for a more detailed discussion.
In addition, the phase is symmetric with respect to $n=3$ owing to the particle-hole symmetry for the spinful system. 
Furthermore, the quantum critical end point is located at $U_{{\rm c}2} \sim 6.8$ and $\lambda \sim 0.003$ for $n =1, 2, 4$, and $5$.


\subsection{$n$--$U$ phase diagrams at $\lambda>0$}
Figures~\ref{fig:n-U-pd}(b), \ref{fig:n-U-pd}(c), and \ref{fig:n-U-pd}(d) show the $n$--$U$ ground-state phase diagrams for $\lambda= 0.003, 0.5, 1.0$, respectively.
As seen in Fig.~\ref{fig:n-U-pd}(b), even for an infinitesimal value of $\lambda$,
the MM-I phase changes into the MM-II phase because $Q$ and $T$ emerge owing to the coupling between $M$ and $Q$/$T$ through $\lambda$ (refer to the discussion in Sec.~V and Fig.~\ref{fig:n4}).

For finite $\lambda$,
the MM-III phase survives up to $\lambda \sim 0.0001$ (not shown).
With further increasing $\lambda$,
the MM-I($b$) phase and the MM-II($a$) phase emerge.
The MM-I($b$) phase is located around $n = 5$. In this phase, the $b$ component $(M_z^b)$ is finite, and the other components are zero. 
Comparing the phase diagrams at $\lambda = 0$ and $0.003$ in Fig.~\ref{fig:n-U-pd}, one can see that the MM-II phase around $n = 5$ is replaced by MM-I($b$).
With an increment in $\lambda$, the MM-I($b$) phase broadens around $n =5$.
In this phase, the two $\jeff$ = 3/2 orbitals are completely filled, and the $\jeff$ = 1/2 orbital remains partially filled. 
As a consequence, the hybridization between $\jeff$ = 3/2 and 1/2 vanishes.

In contrast, the MM-II($a$) phase was located for $n \leq 1$. In this phase, the $a$ components $(M_z^a, Q_u^a$ and $T^{\alpha a}_z)$ are finite, and the other components are zero. The MM-II($a$) phase appears when the energy gap between $\jeff$ = 3/2 and 1/2 is large. 
The $\jeff$ = 1/2 orbital is empty and its contribution is negligible.

\subsection{$U$--$\lambda$ phase diagrams}
\begin{figure}[tbp]
  \begin{center}
   \includegraphics[clip,width=\linewidth]{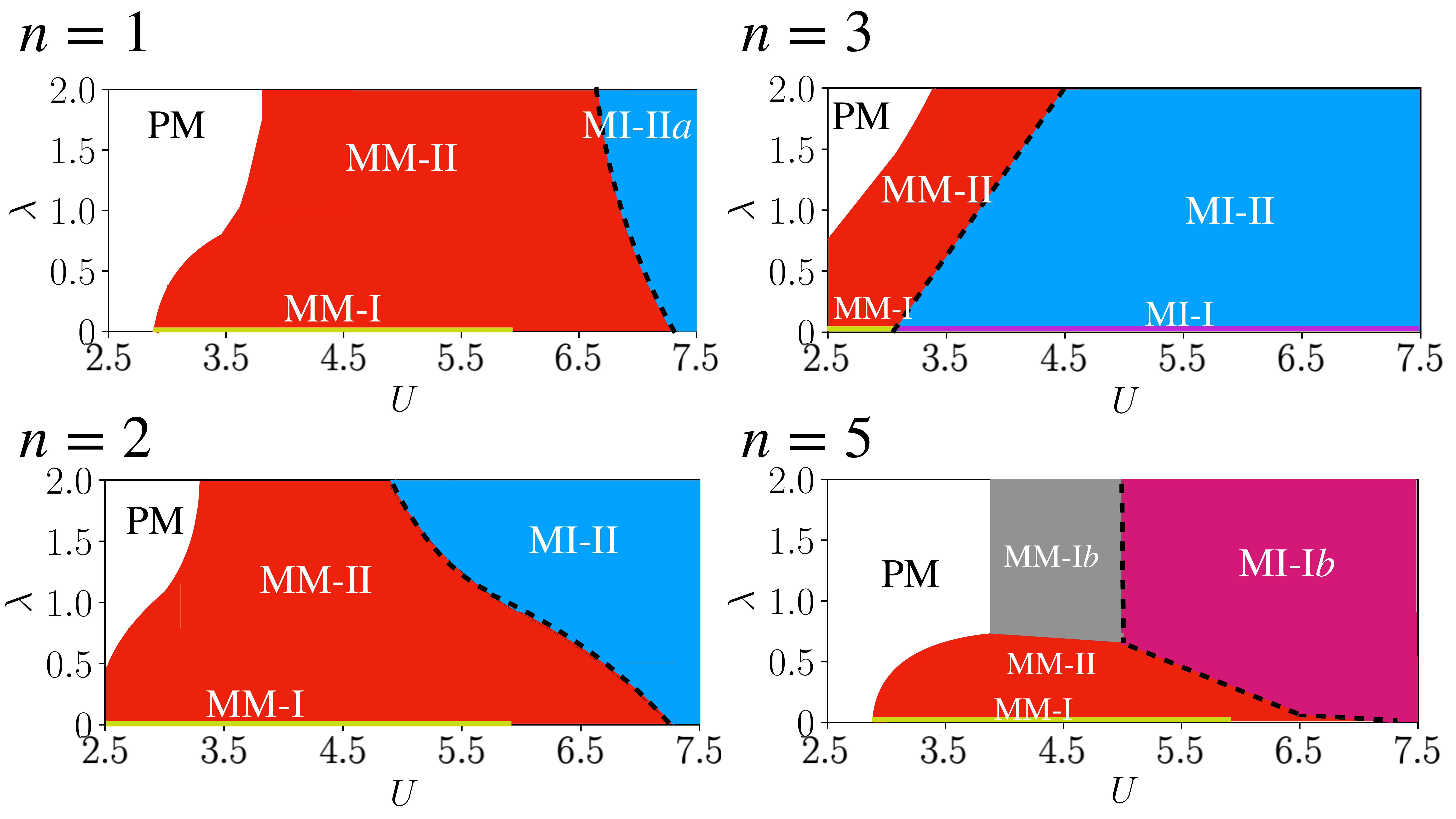}
  \end{center}
  \caption{Phase diagrams computed for $n = 1, 2, 3, 5$.
  The data are the same as those shown in Fig.~\ref{fig:n-U-pd}).
  See the main text and captions of Fig.~\ref{fig:n4}, to define the phases (labels).
  }
  \label{fig:phase1235}
\end{figure}

Figure \ref{fig:phase1235} shows the ground-state $U-\lambda$ phase diagrams for $n = 1, 2, 3$ and $5$ (The phase diagram for $n=4$ was discussed in Sec.~\ref{sec_n4}).

At $n=5$,
the phase boundaries between PM, MM-I$(b)$, and MI-I$(b)$ are vertical for $\lambda \gtrsim 0.7$.
In this regime, the $\jeff$ = 3/2 orbitals are completely filled and the $\jeff$ = 1/2 orbital is half-filled owing to the gap between these two manifolds induced by $\lambda$.
Thus, these transitions can be regarded as phase transitions in the effective single-orbital model of $\jeff$ = 1/2.
This explains why the critical values of $U$ do not depend on $\lambda$.
For $n=3$, the boundary between MM-II and MI-II runs linearly to the upper right.
This may be due to the competition between $\lambda$ and local interactions ($U$, $J_\mathrm{H}$).

The phase diagrams for $n=1$ and $n=2$ are similar in nature. This is because the electrons partially occupy $j_{\rm eff}=3/2$, and the fillings are less than half in both cases. The partially filled orbitals always have magnetic moments regardless of the value of $\lambda$, which is magnetically ordered by interaction effects.
In contrast, at $n=4$, a sufficiently large $\lambda$ suppresses the magnetic order because active moments are absent.

\begin{figure}[htbp]
   \centering
   \includegraphics[clip,width=0.9\linewidth]{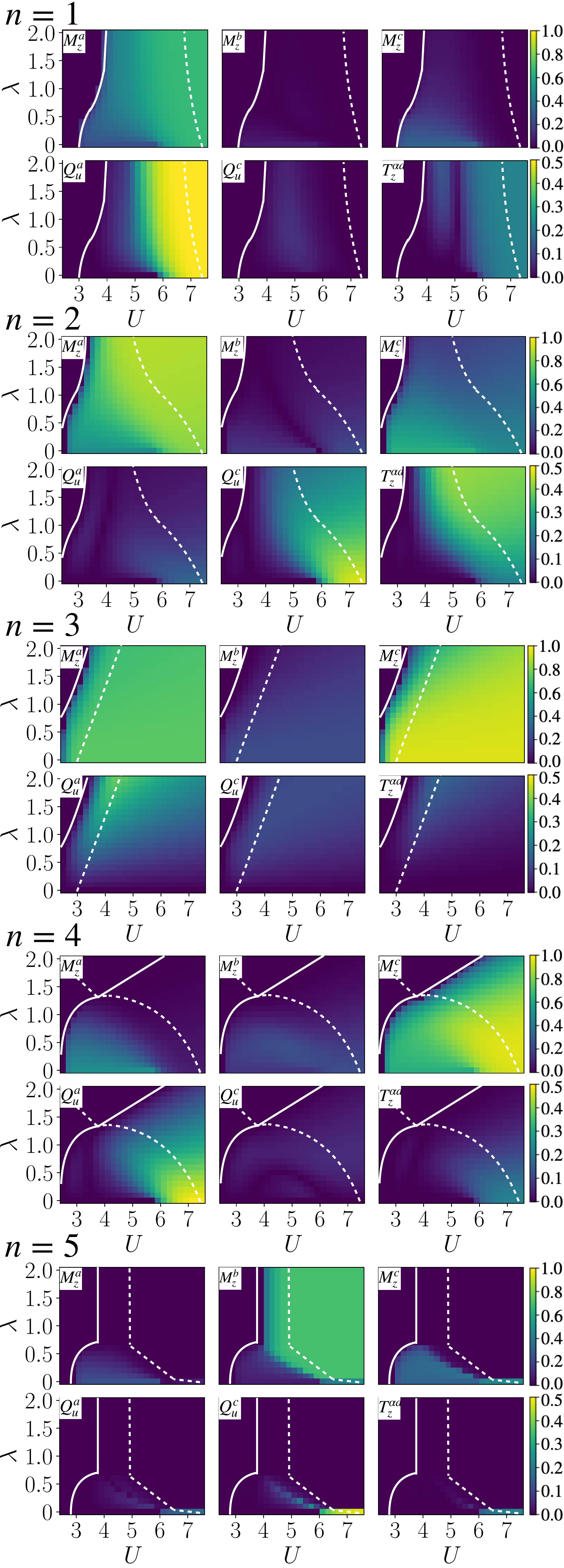}
  \caption{Intensity maps for $n = 1, 2, 3, 4, 5$. The solid and broken lines indicate the magnetic phase transitions and metal-insulator transitions, respectively.
  See Table~\ref{tab:multipolemain} for the definition of the order parameters.
  The superscripts $a, b$, and $c$ denote $\jeff = 3/2$ components, $\jeff = 1/2$ components, and $\jeff = 3/2$--$1/2$ entangled components, respectively.}
  \label{fig:intensity-map}
\end{figure}

Next, the intensity maps of the OPs computed for $n=1, 2, 3, 4$, and $5$ are shown in Fig.~\ref{fig:intensity-map}.
The absolute values of the computed OPs are plotted.
The phase transition lines are marked with white lines.
$\jeff$ = 1/2 diagonal components denoted by $a$ are dominant for $n=5$.
For $n =3$ and $4$, $a$ and $c$ components ($\jeff$ = 3/2--1/2 entangled) are enhanced depending on the balance between $U$ and $\lambda$. As shown in Fig.~\ref{fig:n-U-pd}, a first-order transition with spontaneous symmetry breaking is observed, except for $n = 3$. In addition, as shown in the case of $n =4$ (Sec.~\ref{sec_n4}), higher-order multipoles $Q$ and $T$ are enhanced when crossing the quantum critical line or its crossover line in the direction in which $U$ grows.

The characteristic features of the multipole order parameters at each filling are discussed now.
At $n =1$, symmetry breaking of the orbital enhances $Q_u^a$. Because the electron is mainly stored in the $\jeff$ = 3/2 orbitals regardless of $U$ or $\lambda$, the $a$ component is dominant.
When $U$ is sufficiently large (i.e., $U \geq 7$), all the $b$ and $c$ components become inactive and only the $a$ components remain active. At the same time, the metal-insulator transition occurs, and the MI-II$(a)$ phase appears. A valley of OPs, such as $T_z^{\alpha a}$ at $U =5$, indicates a sign change similar to $Q_u$, as shown in Fig.~\ref{fig:lamb0opold}).

At $n =2$, $Q_u^c$ and $T^{\alpha  a}_z$ are notable.
The former
is enhanced by the symmetry breaking in the orbital sector.
On the other hand, the latter is amplified by the SOC because $\jeff$ = 3/2 and 1/2 are split, and the $a$ component is activated. $T^{\alpha  a}_z$ has a peak at $n = 2$, probably because it belongs to the same irreducible representation $\Gamma_{4u}$ as $M_z$.
 
At $n=3$, either the high-spin state or the low-spin state is stable, depending on $U$ and $\lambda$.
For $U \gg \lambda$, the high-spin state is stable. In this state, the angular momentum $L$ vanishes, and thus $Q$ and $T$ are zero. Owing to this circumstance, the magnetic insulating state without $Q$ and $T$ (MI-I) is realized at $n =3$.
For $U \ll \lambda$, a low-spin state with three electrons in the $\jeff=3/2$ orbitals becomes stable.
In this state,
both, the spin and orbital degrees of freedom remain; thus, higher-order multipoles can be activated.

At $n=4$, as already discussed in Sec.~\ref{sec_n4},
the $c$ component of the magnetic dipole $M_z^c$ is dominant. In the large $\lambda$ region, the paramagnetic insulating (PI) state, which is induced by the strong SOC, appears for $\lambda \geq 1.3$. 

At $n = 5$ and large $\lambda$, the two $\jeff$ = 3/2 orbitals are almost filled, leaving the $\jeff$ = 1/2 orbital half-filled.
Consequently, all the $a$ components are inactive, while the $b$ component $(M_z^b)$ is enhanced. 
At small $U$ and $\lambda$, the $a$ and $c$ components are active because of the hybridization between $\jeff$ = 1/2 and 3/2. 
However, with further increments of $U$ or $\lambda$, the $a$ and $c$ components become inactive and the MM-I($b$) and MI-I($b$) phases emerge.

\section{SUMMARY}\label{sec:summary}
In conclusion, 
the $t_{2{\rm g}}$ Hubbard model using the unrestricted Hartree-Fock approximation with all possible (particle-conserving) local symmetry breaking patterns has been studied.

The main results of this work are twofold.
First, a complete set of multipole order parameters are constructed that can identify entanglements between \jeff = 1/2 and 3/2 manifolds. Second, through extensive Hartree-Fock calculations, the ground-state phase diagrams in the parameter space of the onsite Coulomb repulsion $U$, the strength of the spin-orbit coupling (SOC) $\lambda$, and filling $n$ are systematically investigated. 
It has been determined that this model hosts many non-trivial quantum phases with multipole ordering as well as peculiar phase structures, such as multi-critical points. Furthermore, the intensity maps of the multipolar order parameters for the phase diagrams have been computed.

The results of the present study clearly show that the simple $t_\mathrm{ 2 g }$ model with a semi-circular density of states can host a variety of phenomena.
An interesting future study may be the analyses of the model using a more elaborate method, such as the dynamical mean-field (DMFT) theory, although DMFT calculations suffer from a sign problem. 
Another future direction may be comparisons with $4d$ and $5d$ materials by considering realistic band structures.

\begin{acknowledgments}
N.C. and H.S. were supported by JSPS KAKENHI Grants No. 18H01158 and No. 21H01003, and S.H. by No. 19H01842 and No. 21K03459.
\end{acknowledgments}

\bibliography{multipole0719}

\appendix

\section{Complete basis set with spin-orbit coupling}\label{app_a}

For systems with spin-orbit coupling, the energy level is split into lower $j_{\rm eff}=3/2$ and higher $j_{\rm eff}=1/2$ states.
At the filling $n=4$, the lower orbital is fully occupied, so that no degrees of freedom are left within the $j_{\rm eff}=3/2$ multiplet.
Then, we need to consider the excited $j_{\rm eff}=1/2$ state, which can, in principle, mix through thermal and/or interaction effects.
Such degrees of freedom are called ``excitonic'' in the sense that the transition matrix involves the process from the ground state to the energetically excited states, and are naturally described through the $j_{{\rm eff}}$-off-diagonal matrix elements.
Here the terminology ``$j_{{\rm eff}}$-off-diagonal'' is introduced, in a manner similar to the pairing amplitude in superconductivity which is called the off-diagonal long-range order.
The purpose of this section is to construct a proper complete set of operators to describe the order parameters for multi-orbital systems with spin-orbit coupling.

Let us begin with the total angular momentum
\begin{align}
    \hat{\bm{J}}_{{\rm eff}} &= - \hat{\bm L} + \hat{\bm S}
\end{align}
which is a good quantum number for the local Hamiltonian.
The eigenstates form a complete local basis set $|j_{{\rm eff}},j_{{\rm eff},z}\rangle = (
|\tfrac 3 2, \tfrac 3 2\rangle, 
|\tfrac 3 2, \tfrac 1 2\rangle, 
|\tfrac 3 2, -\tfrac 1 2\rangle,
|\tfrac 3 2, -\tfrac 3 2\rangle, 
|\tfrac 1 2, \tfrac 1 2\rangle, 
|\tfrac 1 2, -\tfrac 1 2\rangle
) $.
For multi-orbital systems, the operators of these states are classified by utilizing the concept of multipole expansion.
The concept of multipoles had been introduced originally for
a description of the local degrees of freedom of $f$ electrons in terms of the total angular momentum $J_{{\rm eff}}$ 
~\cite{Ohkawa1983, Shiina1997, Kusunose2008jpsj, Kuramoto2009, Hayami2018}.

High-rank multipoles are needed that can be introduced as a polynomial form of the $\hat{\bm J}_{{\rm eff}}$ operator \cite{Sato2019}.
However, the $j_{\rm eff}$-off-diagonal components, which are the essential quantities for the order parameters at $n=4$, cannot be described because the amplitude of the total angular momentum $\hat{\bm J}^2_{{\rm eff}}$ is a conserved quantity.
The corresponding set of multipole operators is incomplete.
Hence, one needs to consider an operator that includes the transition between different $j_{\rm eff}$\,s.
Namely, another angular momentum is defined:
\begin{align}
    \bm K = \alpha \hat{\bm L} + \beta \hat{\bm S}
\end{align}
where $\alpha$ and $\beta$ ($\in \mathbb R$) are constants, which in general includes the ``perpendicular'' component with respect to $\hat{\bm J}_{{\rm eff}}$.
This operator also commutes with the local Hamiltonian, but does not commute with $\hat{\bm J}_{{\rm eff}}$.
Multipole operators can be constructed based on the polynomial expressions of $\bm K$ as:
\begin{align}
    N &= 1
    \\
    \bm M &= \bm K
    \\
    Q_{xy} &= \overline{K_x K_y}
    \\
    Q_{yz} &= \overline{K_y K_z}
    \\
    Q_{zx} &= \overline{K_z K_x}
    \\
    Q_{3z^2-r^2} &= 3K_z^2-\bm K^2 \ \ (=Q_u)
    \\
    Q_{x^2-y^2} &= \overline{K_x^2 - K_y^2} \ \ (=Q_v)
    \\
    T_{xyz} &=\overline{K_x K_y K_z}
    \\
    T_{x(5x^2-3r^2)} &= \overline{K_x(5K_x^2 -3\bm K^2)} \ \ (=T_x^\alpha)
    \\
    T_{y(5y^2-3r^2)} &= K_y(5K_y^2 - 3\bm K^2) \ \ (=T_y^\alpha)
    \\
    T_{z(5z^2-3r^2)} &= \overline{K_z(5K_z^2 - 3\bm K^2)} \ \ (=T_z^\alpha)
    \\
    T_{x(y^2-z^2)} &= \overline{K_x(K_y^2 - K_z^2)} \ \ (=T_x^\beta)
    \\
    T_{y(z^2-x^2)} &= \overline{K_y(K_z^2 - K_x^2)} \ \ (=T_y^\beta)
    \\
    T_{z(x^2-y^2)} &= \overline{K_z(K_x^2-K_y^2)} \ \ (=T_z^\beta)
\end{align}
where the overline symmetrizes the expression as $\overline{ABC} = (ABC+ACB+BAC+BCA+CAB+CBA)/3!$, for example.
These operators are referred to as monopole ($N$), dipole ($M_\mu$), quadrupole ($Q_\lambda$), and octupole ($T_\xi$) in accordance with the number of multiplied angular momenta~\cite{Kusunose2020complete,Tamura2020}.
The further high-rank tensors are zero.
$6\times 6=36$ multipoles are expected, but there are only 16 in the above, which is not enough.
The complete matrix basis can be constructed using the local projection operators
$P_{3/2}$ and $P_{1/2}$, each of which singles out the $\jeff=3/2$ and $\jeff=1/2$ components, respectively ($P_{3/2}+P_{1/2} = 1$).
First, the operators in the $\jeff=3/2$ subspace are introduced as
\begin{align}
    N^{\rm 3/2,even} (\equiv N^a) &\propto P_{3/2}
    \\
    M^{\rm 3/2,odd}_\mu  (\equiv M^a_\mu) &\propto P_{3/2} M_\mu P_{3/2}
    \\
    Q^{\rm 3/2,even}_\lambda (\equiv Q^a_\lambda) &\propto P_{3/2} Q_\lambda P_{3/2}
    \\
    T^{\rm 3/2,odd}_\xi (\equiv T^a_\xi) &\propto P_{3/2} T_\xi P_{3/2} = T_\xi
\end{align}
where $\mu, \lambda, \xi$ represent polynomials,
and the ones in $\jeff=1/2$ subspace are defined by
\begin{align}
    N^{\rm 1/2,even} (\equiv M^b_\mu) &\propto P_{1/2}
    \\
    M^{\rm 1/2,odd}_\mu (\equiv M^b_\mu) &\propto P_{1/2} M_\mu P_{1/2}
\end{align}
The ``even'' and ``odd'' symbols represent the sign from the time-reversal operation $\mathscr T = \exp(- i\pi J_{y})\mathscr K$ with the complex conjugation $\mathscr K$.
For examples, one can confirm
\begin{align}
    \mathscr T M^{\rm 3/2,odd}_\mu \mathscr T^{-1} &= - M^{\rm 3/2,odd}_\mu
    \\
    \mathscr T Q^{\rm 3/2,even}_\lambda \mathscr T^{-1} &= + Q^{\rm 3/2,even}_\lambda
\end{align}
and so on.
The quadrupoles and octupoles do not exist for the $\jeff=1/2$ subspace: $P_{1/2}Q_\lambda P_{1/2} = P_{1/2}T_\xi P_{1/2} = 0$.
We used a short-hand notation such as $N^a$ and $M^b$ in the main text (see Tab.~I).

The $j_{{\rm eff}}$-off-diagonal (the symbol ``offd'' assigned) operators are also obtained, which are classified by the time-reversal operation.
The time-reversal odd $j_{\rm eff}$-off-diagonal operators are 
\begin{align}
    M^{\rm offd, odd}_\mu  (\equiv M^c_\mu) &\propto P_{3/2} M_\mu P_{1/2} + P_{1/2} M_\mu P_{3/2}.
    \\
    Q^{\rm offd, odd}_\lambda (\equiv Q^d_\lambda) &\propto -i (P_{3/2} Q_\lambda P_{1/2} -  P_{1/2} Q_\lambda P_{3/2}).
\end{align}
which are magnetic dipoles and magnetic quadrupoles.
In addition, there are time-reversal even ones
\begin{align}
    M^{\rm offd, even}_\mu (\equiv M^d_\mu) &\propto -i (P_{3/2} M_\mu P_{1/2} - P_{1/2} M_\mu P_{3/2}),
    \\
    Q^{\rm offd, even}_\lambda (\equiv Q^c_\lambda) &\propto P_{3/2} Q_\lambda P_{1/2} + P_{1/2} Q_\lambda P_{3/2}.
\end{align}
which are electric dipoles and electric quadrupoles.
In particular, for the electric dipoles, there is another symbolic expression
\begin{align}
    \bm M^{\rm offd, even} \propto \bm L \times \bm S
\end{align}
which is identified as time-reversal even.

Thus, the complete 36 basis matrices have been constructed.
Once the matrix representation of these operators is normalized by the trace of the squared matrices, these matrices do not depend on the constants $\alpha$ and $\beta$.
The simple choice is then $\alpha =\beta = 1$ in the present case.
The above matrices are also orthogonal as
\begin{align}
    {\rm Tr\,} O_\xi O^{\dagger}_{\xi'} = \delta_{\xi\xi'}
\end{align}
where the indices $\xi$ take the 36 types of multipoles.
With these setups, the magnitudes of the expectation values of each multipole operator can be compared and the primary order parameter that is largest in magnitude can be found.

The classification based on the point group is also possible.
The ranks 0,1,2,3 respectively correspond to the monopoles ($N^{\rm 3/2,even}, N^{\rm 1/2,even}$), dipoles ($M^{\rm 3/2,odd}, M^{\rm 1/2,odd}, M^{\rm offd,odd}, M^{\rm offd,even}$), quadrupoles ($Q^{\rm 3/2,even}, Q^{\rm offd,odd}, Q^{\rm offd,even}$), and octupoles ($N^{\rm 3/2,odd}$), respectively, which are defined above.
Each rank is further decomposed based on the cubic harmonics.
The obtained results are summarized in Table~\ref{tab:multipolemain} in the main text.
$j$-parity for the complete classification of the multipole moments is introduced.
This is similar to the case of $sp$ hybridized systems, where the spatial parity is $(-1)^\ell$, giving $+1$ for $s$-electrons ($\ell=0$) and $-1$ for $p$-electrons ($\ell=1$).
In the same manner, a similar parity transformation is considered to distinguish $j_{\rm eff}=3/2$ from $j_{\rm eff}=1/2$, where the total angular momentum differs by 1.
Namely, the transformation matrix is defined using the projection operators as
\begin{align}
    \mathscr P &= P_{3/2} - P_{1/2}
\end{align}
which transforms the wave function as
\begin{align}
    \mathscr P |j_{\rm eff}j_{{\rm eff},z}\rangle &= (-1)^{j_{\rm eff}+1/2} |j_{\rm eff}j_{{\rm eff},z} \rangle
\end{align}
and the transformed operators are, for example, 
\begin{align}
    \mathscr P M_{\mu}^{3/2,\rm odd} \mathscr P^{-1} &= + M_{\mu}^{3/2,\rm odd}
    \\
    \mathscr P M_{\mu}^{\rm offd, odd} \mathscr P^{-1} &= - M_{\mu}^{\rm offd, odd}
\end{align}
The $j$-parity gives $-1$ for the $\jeff$-off-diagonal components.
Thus, every multipole is uniquely classified in terms of rank, time reversal, and $j$parity.
Note that all the multipoles in this study are even under real spatial parity transformation because the $d$ electrons are considered.
It is also noted that the concept of $j$-parity is introduced for the classification of the multipole moments, and the interacting Hamiltonian is not invariant under this transformation.
This is why $j_{\rm eff}$-off-diagonal multipoles mix in general, as shown in the numerical results.

The multipole expansion is regarded as the choice of a set of basis matrices and is not unique.
Our multipole basis based on the total angular momentum is different from the previously proposed ones \cite{Wang2017Eu2,Kusunose2020complete} in that $j_{\rm eff}$-diagonal and $j_{\rm eff}$-off-diagonal components are classified.
If we restrict ourselves to the $\jeff=3/2$ diagonal subspace, the matrices are the same as those used in Ref.~\cite{Tamura2020}.

\section{Symmetry in orbital space}\label{appendix:symm}
Here a comment on the symmetry in orbital space is made.
In the case without spin-orbit coupling, the Hamiltonian with the hopping term and Slater-Kanamori interaction has SO(3) symmetry.
Namely, the Hamiltonian is invariant under the transformation
\begin{align}
    c_{i\alpha \sigma} \longrightarrow \sum_\beta V_{\alpha \beta} c_{i\beta \sigma}
\end{align}
where $V_{\alpha\beta} \in \mathbb R$ is an orthogonal $3\times 3$ matrix.

If the full-spin polarized situation is considered, as in phase MM(II) in Fig.~8, then, only the $\sigma=\uparrow$ components need to be looked at.
Then, the symmetry in the orbital space is elevated to SU(3), where the transformation matrix $V$ is unitary with complex variables.
This emergent symmetry is broken once the down spin components are mixed.
In addition to the above spin-polarized situation, when the spin-orbit coupling is turned on, we have an additional term $\propto L_z S_z$, which acts as the magnetic field in the orbital space.

It is shown that a full-spin polarization significantly simplifies the self-consistent equations.
The Hamiltonian is
\begin{align}
    \mathscr H &= \sum_{\bm k\alpha} (\varepsilon_{\bm k}-\mu) c^\dagger_{\bm k\alpha \uparrow}
    c_{\bm k\alpha \uparrow}
    + \frac{\lambda}{2} \sum_{i\alpha\beta} \ell^z_{\alpha\beta}
    c^\dagger_{i\alpha \uparrow}
    c_{i\beta \uparrow}
    \nonumber \\
    &+ \frac{U-J}{2} \sum_{\alpha\beta} 
    c^\dagger_{i\alpha \uparrow}
    c_{i\alpha \uparrow}
    c^\dagger_{i\beta \uparrow}
    c_{i\beta \uparrow}
\end{align}
where $\alpha,\beta = xy,yz,zx$.
It is notable that the Hamiltonian has a particle-hole symmetry within the $\uparrow$-spin sector.
The $\alpha$-basis is changed to the $\ell_z(=m)$basis by a unitary transformation ($m=0,\pm 1$).
Then the Hamiltonian becomes
\begin{align}
    \mathscr H &= \sum_{\bm km} (\varepsilon_{\bm k}-\mu) c^\dagger_{\bm km \uparrow}
    c_{\bm km \uparrow}
    + \frac{\lambda}{2} \sum_{im} m
    c^\dagger_{im \uparrow}
    c_{im \uparrow}
    \nonumber \\
    &+ \frac{U-J}{2} \sum_{mm'} 
    c^\dagger_{im \uparrow}
    c_{im \uparrow}
    c^\dagger_{im' \uparrow}
    c_{im' \uparrow}
\end{align}
The density-type mean-field is assumed and the self-consistent equations obtained, as follows :
\begin{align}
    \Delta_m
    &= 
(U-J) \sum_{m'\neq m}
F( \mu - \Delta_{m'} - \frac{\lambda}{2} m' )
\\
n &= \sum_m F( \mu - \Delta_{m} - \frac{\lambda}{2} m )
\end{align}
at zero temperature, where
\begin{align}
    F(\varepsilon) &= \int_{-\infty}^{\varepsilon} D(\varepsilon') d\varepsilon'
\end{align}
is the integrated density of states.
The above argument relies only on the spin-polarized situation and is applicable to cases with general filling, interaction, and small spin-orbit coupling.
It is confirmed that the solution of these simplified equations is consistent with the ones discussed in the main text inside the magnetic phases at small $\lambda$.

\end{document}